\documentclass[twocolumn,showpacs,superscriptaddress,10pt]{revtex4-1} 

\usepackage[utf8]{inputenc}
\usepackage{hyperref}
\usepackage{epsfig}
\usepackage{graphicx}
\usepackage{amsmath}
\usepackage{amssymb}
\usepackage{color}
\graphicspath{{Images/}}

\begin{document}

\title{
Holographic laser Doppler imaging of microvascular blood flow
}

\author{C. Magnain}

\affiliation{Institut Langevin. Centre National de la Recherche Scientifique (CNRS) UMR 7587, Institut National de la Sant\'e et de la Recherche M\'edicale (INSERM) U 979, Universit\'e Pierre et Marie Curie (UPMC), Universit\'e Paris Diderot. Ecole Sup\'erieure de Physique et de Chimie Industrielles - 10 rue Vauquelin. 75005 Paris, France}

\author{A. Castel}

\affiliation{Institut Langevin. Centre National de la Recherche Scientifique (CNRS) UMR 7587, Institut National de la Sant\'e et de la Recherche M\'edicale (INSERM) U 979, Universit\'e Pierre et Marie Curie (UPMC), Universit\'e Paris Diderot. Ecole Sup\'erieure de Physique et de Chimie Industrielles - 10 rue Vauquelin. 75005 Paris, France}

\author{T. Boucneau}

\affiliation{Institut Langevin. Centre National de la Recherche Scientifique (CNRS) UMR 7587, Institut National de la Sant\'e et de la Recherche M\'edicale (INSERM) U 979, Universit\'e Pierre et Marie Curie (UPMC), Universit\'e Paris Diderot. Ecole Sup\'erieure de Physique et de Chimie Industrielles - 10 rue Vauquelin. 75005 Paris, France}

\author{M. Simonutti}

\affiliation{Institut de la Vision, INSERM UMR-S 968. CNRS UMR 7210. UPMC. 17 rue Moreau, 75012 Paris. France}

\author{I. Ferezou}

\affiliation{Brain Plasticity Unit, CNRS UMR 8249, ESPCI ParisTech, 10 Rue Vauquelin, 75005 Paris, France}

\author{A. Rancillac}

\affiliation{Brain Plasticity Unit, CNRS UMR 8249, ESPCI ParisTech, 10 Rue Vauquelin, 75005 Paris, France}

\author{T. Vitalis}

\affiliation{Brain Plasticity Unit, CNRS UMR 8249, ESPCI ParisTech, 10 Rue Vauquelin, 75005 Paris, France}

\author{J. A. Sahel}

\affiliation{Institut de la Vision, INSERM UMR-S 968. CNRS UMR 7210. UPMC. 17 rue Moreau, 75012 Paris. France}

\author{M. Paques}

\affiliation{Institut de la Vision, INSERM UMR-S 968. CNRS UMR 7210. UPMC. 17 rue Moreau, 75012 Paris. France}

\author{M. Atlan}

\affiliation{Institut Langevin. Centre National de la Recherche Scientifique (CNRS) UMR 7587, Institut National de la Sant\'e et de la Recherche M\'edicale (INSERM) U 979, Universit\'e Pierre et Marie Curie (UPMC), Universit\'e Paris Diderot. Ecole Sup\'erieure de Physique et de Chimie Industrielles - 10 rue Vauquelin. 75005 Paris, France}

\begin{abstract}

We report on local superficial blood flow monitoring in biological tissue from laser Doppler holographic imaging. In time-averaging recording conditions, holography acts as a narrowband bandpass filter, which, combined with a frequency-shifted reference beam, permits frequency-selective imaging in the radiofrequency range. These Doppler images are acquired with an off-axis Mach–Zehnder interferometer. Microvascular hemodynamic components mapping is performed in the cerebral cortex of the mouse and the eye fundus of the rat with near-infrared laser light without any exogenous marker. These measures are made from a basic inverse-method analysis of local first-order optical fluctuation spectra at low radiofrequencies, from 0 Hz to 100 kHz. Local quadratic velocity is derived from Doppler broadenings induced by fluid flows, with elementary diffusing wave spectroscopy formalism in backscattering configuration. We demonstrate quadratic mean velocity assessment in the 0.1-10 mm/s range in vitro and imaging of superficial blood perfusion with a spatial resolution of about 10 micrometers in rodent models of cortical and retinal blood flow.

\end{abstract}

\maketitle 

\section{Introduction}

\subsection{Motivations}

The role of the microcirculation is increasingly being recognized in the pathophysiology of cardiovascular diseases; eg. hypertension~\cite{Gooding2010, StruijkerBoudier2012, Book_StruijkerBoudier2012} and diabetes~\cite{Rendell1989, Schlager2012, VasGreen2012}. In particular, assessing retinal blood flow can be potentially useful for understanding diabetic retinopathies~\cite{Grunwald1993}, and study the relationships between vascularization and glaucoma~\cite{Koseki2008, Venkataraman2010, Sugiyama2011, Schmidl2012, HwangKonduru2012}. Moreover, extensive use of optical methods is made for monitoring skin microvascular endothelial (dys)function~\cite{Kubli2000, CracowskiMinson2006, Holowatz2008, KhanPatterson2008, Turner2008, VanHerptDraijer2010}. Prevalent optical techniques to monitor microvascular blood flow in clinical studies are laser Doppler probes and spatial speckle contrast imaging. The former is characterized by its high temporal resolution and the latter enables wide-field imaging of superficial microvascular networks.  Superficial blood flow monitoring require sensitivity in low light, high temporal resolution, high spatial resolution, and the ability to perform quantitative flow measurements. Most current techniques are limited in their spatial resolution or temporal resolution or both. Hence wide-field optical imaging techniques using laser light and sensor arrays to probe local dynamics with potentially high spatial and temporal resolution are attracting attention for the measurement of blood flow~\cite{HumeauHeurtier2013}.

\subsection{Relationship between local motion and optical fluctuations}

The radiofrequency (RF) spectrum of dynamic light fluctuations is affected by microvascular hemodynamics and hence a subject of great interest for blood flow imaging applications. The laser Doppler technique measures Doppler-shifts and broadenings of quasi-elastically scattered light. Depending on the detection configuration, single scattering or multiple scattering can be targeted to yield Doppler spectra. The observation and interpretation of Doppler broadening of a scattered laser light beam by a fluid in motion in vitro~\cite{YehCummins1964} and in vivo~\cite{Stern1975} has led to the physical modeling of laser Doppler velocimetry~\cite{BonnerNossal1981, HeckmeierSkipetrov1997}. Mapping retinal hemodynamics \textit{in vivo} has many biomedical applications such as diagnosis of retinal microvasculature disorders~\cite{Feke1978, Schmetterer1999, SatoFeke2006}. Optical instrumentation is well adapted to non-invasive retinal blood flow imaging, because the eye fundus vascular tree is visible through the cornea and is thus potentially accessible for light imaging. The Doppler shift of a monochromatic optical radiation scattered by a moving target is the scalar product of the optical momentum transfer with the target velocity. Doppler shifts are cumulative. In the case of multiple scattering, Algebraic Doppler shifts add-up throughout each optical path, and the broadening of the backscattered radiation still carries a highly valuable information, for velocity assessment~\cite{Riva1972, BonnerNossal1981}. Under coherent illumination, the local motion of scatterers can be probed either by the analysis of local spatial contrast, the local temporal fluctuations, or the local RF Doppler spectrum of the speckle pattern~\cite{Briers1996}. Single point laser Doppler detection schemes in heterodyne~\cite{Riva1979, Michelson1996, Ferguson2004, RajanVarghese2009} or self-mixing~\cite{NilssonTenland1980, DeMulVanSpijker1984, Bosch2001, NikolicHicks2013} configurations are prevalent in practical detection of blood flow from wideband temporal optical fluctuations analysis, for superficial blood flow sensing. The design of depth-resolved Doppler-contrast optical coherence tomography schemes~\cite{Izatt1997, LeitgebSchmetterer2003, AnWang2008, AnLiLan2013} is also an active field of research with applications in vascular imaging of the posterior segment of the eye. In this article, we limit the description of state-of-the-art detection schemes based on sensor arrays.

\subsection{Spatial speckle contrast analysis}

Spatial intensity fluctuation analysis techniques~\cite{Draijer2009R, BoasDunn2010, Dunn2012, Basak2012, ValdesVarma2014} can be referred to as time-averaged speckle contrast imaging. These techniques have been adopted over the past decades for imaging of blood flow dynamics in real-time. Their rapid adoption for physiological studies is due to to the ability to quantify blood flow changes with good spatial and temporal resolution. Local motion imaging by the analysis of spatial contrast of laser speckle is a simple and robust method to achieve full-field imaging~\cite{FercherBriers1981, Briers1995, Briers2001}. It is a widely used imaging technique to image blood flow in vivo, primarily due to advantages like ease of instrumentation, reproducibility~\cite{Roustit2010}, and low cost~\cite{RichardsKazmi2013}. It relies on local speckle contrast in time-averaged recording conditions to assess local motion of light scatterers. Motion of scattering particles, like the red blood cells, causes spatial and temporal blurring of the speckle pattern. In speckle contrast imaging, the interference pattern is recorded with a camera, and the blurring of the pattern is quantified to obtain a measure of relative change in flow. Spatial contrast analysis techniques improved to assessment of quantitative relative changes in local blood perfusion with exposure-control of the recorded frames~\cite{Bandyopadhyay2005, ParthasarathyTom2008} and consideration of the non-fluctuating light component in contrast analysis models~\cite{ZakharovVolker2006}. Speckle contrast analysis schemes were used to generate images of blood flow in the rat retina in traditional configuration~\cite{ChengDuong2007, SriencKurthNelson2010}, and via an endoscope~\cite{PonticorvoCardenas2013}. Speckle contrast imaging is now being routinely applied to cerebral blood flow assessment, providing valuable vascular perfusion information in vivo~\cite{DunnBolay2001, DunnDevor2003, DunnDevor2005, DuncanKirkpatrick2008, Zakharov2009}; in particular with the advent of multi-exposure speckle imaging~\cite{ParthasarathyTom2008, ParthasarathyKazmi2010}. Recently, the technique was used with success to monitor cerebral blood flow during neurosurgery in the human brain~\cite{ParthasarathyWeber2010, Klijn2013}. 

\subsection{Temporal speckle contrast analysis with sensor arrays}

Full-field laser Doppler imaging techniques from time-domain optical intensity fluctuations measurements on sensor arrays can enable monitoring of blood perfusion changes~\cite{Serov2002, Serov2005, Serov2005, ChengYan2008, Draijer2009, WangMoyer2012, ZengWangFeng2013, WangZengLiangFeng2013}. In these approaches, laser Doppler signal processing is performed off-chip, which means that large digital data transfers have to be handled from the sensor array to the processing unit of a computer. Recently, image-plane laser Doppler recordings with a high throughput complementary metal–oxide–semiconductor (CMOS) camera in conjunction with short-time discrete Fourier transform calculations by a field programmable gate array (FPGA) reportedly enabled continuous monitoring of blood perfusion in the mm/s range. Full-field flow maps of 480 $\times$ 480 pixels were rendered at a rate of 14 Hz, obtained from image recordings at a frame rate of 14.9 kHz~\cite{Leutenegger2011}. However, cameras capable of frame rates of up to several kilohertz and data transfer rates still larger than the commercially available ones are a requirement in this approach. Locally processing the laser Doppler signal within the sensor array can prevent large data throughputs off-chip, which is presently a technological bottleneck. To circumvent data transfer issues, fully-integrated CMOS sensor arrays with Doppler signal processing were developed~\cite{GuHayesGill2008}, which have the advantage of enabling not only digital but also analog signal processing, which permits capacitive coupling to cancel DC contributions and low-pass filtering of the analog signal to process. Such sensors were demonstrated to achieve 64 $\times$ 64 pixel blood flow images at 1 frame per second~\cite{HeNguyen2012, NguyenHayesGill2011, HeNguyen2013}, with a sampling bandwidth of 40 kHz. On-chip laser Doppler signal processing is one of the most promising approaches for the temporal analysis of optical fluctuations for blood flow imaging. It alleviates the technical issue of unmet data throughput required in time-resolved parallel sensing with sensor arrays.

\subsection{Laser Doppler imaging by holography}

Early experimental demonstrations of holographic ophthalmoscopy of the eye fundus in vivo~\cite{Calkins1970, Wiggins1972, Rosen1975, OhzuKawara1979, Tokuda1980} have shown the potential of this modality for retinal vessels imaging, before the advent of digital acquisition schemes with sensor arrays. A convenient way to perform optical heterodyne detection with video frame rate sensor arrays is to have recourse to holography. For instance, blood flow contrasts were imaged by holography in the cerebral cortex~\cite{AtlanGrossVitalis2006} and in the eye fundus~\cite{SimonuttiPaquesSahel2010} of rodents with a standard camera. Time-averaged holography with a frequency-shifted reference beam was shown to enable narrow band detection and imaging of local Doppler broadenings~\cite{AtlanGrossLeng2006, AtlanGross2006, VerrierAlexandreGross2014}. Blood flow contrasts were also successfully rendered with wideband detection of optical fluctuations in holographic configuration~\cite{AtlanGrossVitalis2008}. Laser Doppler holography enables narrowband optical heterodyne detection with the benefit of canceling laser amplitude noise, which can be efficiently filtered spatially in off-axis recording configuration~\cite{AtlanGross2007JOSAA} at the price of a reduction of the available spatial bandwidth. Hence holographic schemes have a major advantage in comparison with direct image recording schemes, especially for narrowband Doppler detection with video-rate cameras.


\subsection{Presented results}

We report on quantitative fluid flow assessment in vitro and in vivo from holographic interferometry, which enables wide-field imaging of a Doppler-shifted radiation with an array detector. Time-averaged holography acts as a narrowband bandpass filter, described in section~\ref{sect_NarrowbandHeterodyneDetection}, which permits frequency-selective Doppler imaging in the radiofrequency range, in combination with a frequency-shifted reference beam. These Doppler images are acquired with an off-axis Mach-Zehnder interferometer, in reflection configuration~\cite{AtlanGrossVitalis2006, SimonuttiPaquesSahel2010}. The signal obtained with narrowband, frequency-tunable holographic interferometry is the first-order power spectrum of the optical field scattered by the object, recorded in heterodyne configuration~\cite{BoasYodh1997}. Multiply-scattered light yields Doppler spectra from which the directional information is lost because of wave vector randomization during scattering. Nevertheless, it has the advantage of providing spatially-resolved wide-field hemodynamic contrast. Our strategy is to use an inverse-problem approach to derive velocity maps of blood flow from optically-acquired Doppler maps; we make use of elementary results from the diffusing-wave spectroscopy formalism~\cite{Pine1988} in backscattering configuration in order to retrieve local root mean square velocities~\cite{GandjbakhcheNossal1993}. The measured heterodyne signal described in section~\ref{sect_NarrowbandHeterodyneDetection} is modeled with the diffusing-wave spectroscopy formalism for flowing and diffusing scatterers in sections~\ref{sect_InVitro}, \ref{sect_DopplerBrain}, and \ref{sect_DopplerEye}. In vitro flow measurements in a calibrated sample are reported in section~\ref{sect_InVitro}. In vivo microvascular blood flow mapping in the cerebral cortex of the mouse and the eye fundus of the rat are reported in section~\ref{sect_DopplerBrain}, and section~\ref{sect_DopplerEye}, respectively.

\section{Narrowband heterodyne detection of light by frequency-tunable holography}\label{sect_NarrowbandHeterodyneDetection}

\subsection{Optical fields}

In this section, we use a scalar, deterministic representation of light to highlight the temporal filtering features of time-averaged heterodyne holography. The single-frequency laser light wave can be modeled by a complex number $E$, which represents the oscillating electric field. This field oscillates at the angular frequency $\omega_{\rm L}$. Polarization properties are neglected in this representation. The amplitude and phase of the field are the absolute value and angle of $E$. The optical field illuminating the object under investigation is of the form $E = E_0 \exp \left( i \omega_{\rm L} t \right)$, where $E_0$ is a complex constant and $i$ is the imaginary unit. The probe optical field, backscattered by the preparation under investigation is noted
\begin{equation}\label{eq_E_field}
E(t) = {\cal E} (t) \exp \left( i \omega_{\rm L} t \right)
\end{equation}
where the envelope ${\cal E} (t)$ carries the RF temporal fluctuations of the field. It is decomposed in discrete frequency components 
\begin{equation}\label{eq_calE_decomposition}
{\cal E} (t) = \sum _{n} \tilde{{\cal E}}(\omega_n) \exp \left( i \omega_n t \right)
\end{equation}
With these notations, the discrete first-order RF power spectrum of the object light is 
\begin{equation}\label{eq_SpectrumDefinition}
s_1(\omega_n) = | \tilde{{\cal E}}(\omega_n) | ^2
\end{equation}
The optical local oscillator (LO) field is a monochromatic wave, coherent with the illumination beam, frequency-shifted by $\Delta \omega$
\begin{equation}\label{eq_ELO_field}
E_{\rm LO}(t) = \tilde{{\cal E}}_{\rm LO} \exp \left( i \Delta \omega t \right) \exp \left( i \omega_{\rm L} t \right) 
\end{equation}
The irradiance of the light wave, in terms of its electric field $E$, is $|E|^2 c \epsilon_0 / 2$, where $\epsilon_0$ is  the permittivity and $c$ is the speed of light in vacuum, with a relative permittivity and a magnetic permeability of 1. At time $t$, the optical power (irradiance integrated over the surface $d^2$ of a pixel) integrated during the exposure time $\tau_{\rm E}$ by the square-law sensor array yields the recorded interferogram
\begin{equation}\label{eq_I_CameraDetection}
I(t) = \frac{1}{2} \epsilon_0 c d^2 \eta \int_{-\tau_{\rm E}/2}^{\tau_{\rm E}/2}  | E( t + \tau ) + E_{\rm LO}( t + \tau ) |^2 {\rm d}\tau
\end{equation}
where $\eta$ is the quantum efficiency of a pixel.

\subsection{Bandpass filtering properties of time-averaged holograms}

The object field $E$ beats against the LO field $E_{\rm LO}$ and yields the following cross-terms contribution in an off-axis region of the spatial frequency spectrum of the interferograms~\cite{AtlanGross2007JOSAA}
\begin{equation}\label{eq_H_t}
H(t) =  \frac{1}{2} \epsilon_0 c d^2 \eta \int_{-\tau_{\rm E}/2}^{\tau_{\rm E}/2}  E( t + \tau )  E^*_{\rm LO}( t + \tau ) {\rm d}\tau
\end{equation}
To calculate an image, the recorded interferograms have to be demodulated spatially. To this end, numerical image rendering in the object plane is performed with a discrete Fresnel transform involving one fast Fourier transform of the recorded interferograms~\cite{SchnarsJuptner1994, Schnars2002, KimYuMann2006, PicartLeval2008, VerrierAtlan2011}. Once this operation is performed, Eq.~\ref{eq_H_t} represents quantities back-propagated to the object plane. The integration of the interference pattern over the exposure time acts as a temporal frequency filter. In the Fourier domain, it corresponds to a low bandpass filter centered at $\Delta \omega$ and whose bandwidth is $1/\tau_{\rm E}$.\\
\begin{equation}\label{eq_H_t_dvlp}
H(t) = C \sum _{n} \tilde{{\cal E}}(\omega_n) \tilde{{\cal E}}^*_{\rm LO} \frac{\sin \left( \Delta \omega_n \tau_{\rm E} /2 \right)}{\Delta \omega_n \tau_{\rm E}} \exp \left( i \Delta \omega_n t \right)
\end{equation}
where $C = \epsilon_0 c d^2 \eta \tau_{\rm E}$ and $\Delta \omega_n = \omega_n - \Delta \omega$ is the angular frequency detuning between the local oscillator and the $n$-th spectral component. Two-phase temporal signal demodulation is performed by calculating the squared magnitude of the difference of two consecutive holograms in order to filter-off stray LO intensity fluctuations from the interferometric cross-terms. The resulting signal is 
\begin{equation}\label{eq_2PhaseDemodulation}
\left| \left( H(t + 2 \pi/\omega_{\rm S}) - H(t) \right) \right|^2 = S^2(t)
\end{equation}
Optical phase fluctuations of the probe field $E$ are considered a stationary process at the time scale of the acquisition of a few frames, such that the signal temporal dependency can be dropped : $S^2(t) \equiv S^2$. Using Eq.~\ref{eq_H_t_dvlp}, Eq.~\ref{eq_2PhaseDemodulation} can be rewritten as
\begin{equation}\label{eq_2PhaseDemodulation2}
S^2 = \left| 2C \tilde{{\cal E}}^*_{\rm LO} \sum _{n} \tilde{{\cal E}}(\omega_n)  \, b(\omega_n - \Delta \omega) \right|^2
\end{equation}
The right member of equation~\ref{eq_2PhaseDemodulation2} is proportional to the squared magnitude of the discrete convolution of $\tilde{{\cal E}}$ with the response function $b$, assessed at the frequency $\Delta \omega$. Besides a constant phase factor, this response is~\cite{AtlanDesbiolles2010}
\begin{equation}\label{eq_fPSF}
    b(\omega)  = \sin \left( {\pi \omega}/{\omega _{\rm S}} \right)\sin \left( \omega \tau_{\rm E}/2 \right)/(\omega \tau_{\rm E}) 
\end{equation}
The response for the probe field's irradiance $B(\omega) = |b(\omega)|^2$ is a bandpass filter of width $\sim 1/\tau_{\rm E}$. Its line shape is reported in Fig.~\ref{fig_TubeSpectra} (dotted line) in typical recording conditions, for $\tau_{\rm E}$ = 30 ms and $\omega_{\rm S}/(2\pi)$ = 12 Hz. 

\subsection{First-order spectrum assessment}

If $1/\tau_{\rm E}$ is much smaller than the typical cutoff frequency of the RF Doppler broadening of the probe beam, we can consider that only the component of the discrete spectral decomposition of the object field at $\Delta \omega$ is measured 
\begin{equation}\label{eq_signal}
S^2 \propto |\tilde{\cal E} (\Delta \omega)|^2 \, |\tilde{\cal E}_{\rm LO}|^2
\end{equation}
The average number of probe photons reaching the pixels of size $d$ during the measurement time $\tau_{\rm E}$ and collected by the filter $B$ tuned on $\Delta \omega$, is noted $n(\Delta \omega)$; it satisfies $n(\Delta \omega) \hbar \omega_{\rm L} \approx  |\tilde{\cal E} (\Delta \omega)|^2 \epsilon_0 c d^2 \tau_{\rm E}/ 2$. The average number of photons of the local oscillator satisfies $n_{\rm LO} \hbar \omega_{\rm L} \approx |\tilde{\cal E}_{\rm LO}|^2 \epsilon_0 c d^2 \tau_{\rm E}/ 2$. In the off-axis region of the hologram, where the object-LO cross-term is separated from other interferometric contributions, the shot-noise is the dominating noise, even in low light~\cite{GrossAtlan2007}. The shot-noise $N$, in high heterodyne gain regime, scales up linearly with $\eta$ and the average number of photons in the LO channel~\cite{WhitmanKorpel1969, UedaMiida1976, Monchalin1985, WagnerSpicer1987, RoyerDieulesaint1986, GrossAtlan2007}. It satisfies 
\begin{equation}\label{eq_noise}
N^2 \propto |\tilde{\cal E}_{\rm LO}|^2
\end{equation}
The domain where this quantity is averaged in practice is shown in Fig.~\ref{fig_TubeHolograms}. From Eq.~\ref{eq_signal}, Eq.~\ref{eq_noise}, and Eq.~\ref{eq_SpectrumDefinition}, a robust narrowband measurement of the RF power spectrum can be performed by forming the following ratio, which does not depend on the LO power 
\begin{equation}\label{eq_SNR}
S^2 / N^2 \propto s_1 (\Delta\omega)
\end{equation}
Where $\Delta \omega$ is the detuning angular frequency of the optical local oscillator. In the case of two-phase demodulation~\cite{AtlanGross2007JOSAA, AtlanDesbiolles2010} (Eq.~\ref{eq_2PhaseDemodulation}), the detection frequency is equal to the detuning frequency of the local oscillator.

\section{Experimental setup}\label{sect_ExperimentalSetup}

\begin{figure}[b]
\centering
\includegraphics[width = \linewidth]{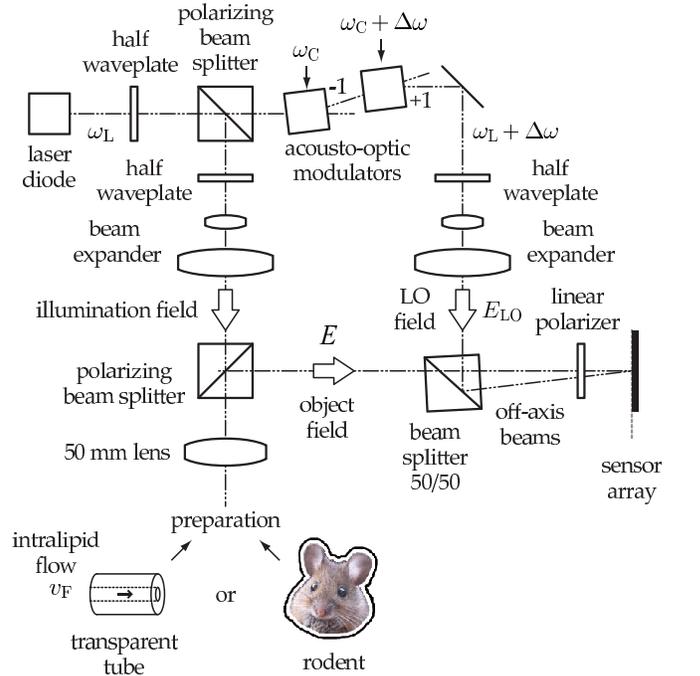}
\caption{Sketch of the optical arrangement : Mach-Zehnder holographic interferometer. The main laser beam of a near infrared laser diode is split into two channels. In the object channel, the optical field $E$ is backscattered by the preparation. In vitro measurements were performed in a transparent tube wherein a controlled flow is induced. For in vivo measurements, the preparation was a cortical window or the eye fundus of rodents. Local motion induces RF broadenings of the optical radiation. These broadenings are quantitatively analysed by narrowband optical detection, at the detuning frequency of the reference channel. In the reference channel, the optical field $E_{\rm LO}$ is frequency-shifted by two acousto-optic modulators from which alternate diffraction orders ($\pm1$) are selected, yielding an optical LO the form of Eq. \eqref{eq_ELO_field}. The sensor array of the camera records the interference pattern $I$ of both optical fields beating against each other, in time-averaging conditions, at a frame rate $\omega_{\rm S}$ and exposure time $\tau_{\rm E}$, which define the narrowband filter of the detection (Eq. \eqref{eq_fPSF}). Images of the preparation are calculated with a standard holographic rendering algorithm involving a numerical Fresnel transform.}\label{fig_setup}
\end{figure}

The experimental laser Doppler imaging scheme used for this study is sketched in Fig.~\ref{fig_setup}. It is based on a heterodyne holographic arrangement previously reported~\cite{AtlanGross2006, AtlanGross2007JOSAA, AtlanGrossVitalis2006, SimonuttiPaquesSahel2010}, which consists of a modified Mach-Zehnder optical interferometer designed for off-axis and frequency-shifting holographic imaging in time-averaged recording conditions~\cite{PicartLeval2003}. In vitro and in vivo flows are investigated. The in vitro preparation, reported in section~\ref{sect_InVitro}, consists of a transparent tube of 1 mm diameter in which an emulsion is injected with a known average velocity. In vivo flow studies of cerebral and retinal blood flow are reported in section~\ref{sect_DopplerBrain}, and section~\ref{sect_DopplerEye}. In all these experiments, a continuous laser diode (Mitsubishi ML120G21) emits a monochromatic radiation of wavelength $\lambda = 785 \, \rm nm$ with an average power of 80 mW. The detection consists of wide-field illumination and collection of the cross-polarized backscattered light from the brain along the same optical axis. It is achieved with a polarizing beam splitter, in order to select photons which have undergone at least a few scattering events~\cite{Schmitt1992}. This approach is chosen to increase the relative weight of multiply scattered Doppler-shifted photons with respect to photons scattered once. The incident light beam is expanded to form a plane wave and diaphragmed by a $\sim 5 \,{\rm mm} \times 5 \,{\rm mm}$ pupil. Its polarization angle is tuned with a half-wave plate to set the illumination power to $\sim$ 1 mW. In the reference arm, an attenuator, a half-wave plate, and a beam expander are used to control the LO beam power, polarization angle, and to ensure a flat illumination of the detector. Two acousto-optic modulators (AA Opto Electronic), driven with phase-locked signals at $\omega_{\rm C}$ and $\omega_{\rm C} + \Delta \omega$, are used to shift the optical frequency of the laser beam from $\omega_{\rm L}$ to $\omega_{\rm L} + \Delta\omega$, to form a frequency-shifted optical local oscillator. The carrier frequency $\omega_{\rm C}/(2\pi)$ is set at the peak response of the acousto-optic modulators, at 80 MHz. The backscattered field $E$ is mixed with the LO field $E_{\rm LO}$ with a non-polarizing beam splitter cube. The interference pattern $I(t)$ is measured at time $t$ by a charge-coupled device array detector (PCO pixelfly QE camera, $1394 \times 1024$ square pixels of $d$ = 6.7 $\mu \rm m$, frame rate $\omega_{\rm S} /(2\pi) = 12 \, \rm Hz$, exposure time $\tau_{\rm E} \simeq 30 \, \rm ms$, dynamic range 12 bit), set at a distance $\Delta z \simeq 69 \, \rm cm$ from the object plane. A small angular tilt $\sim 1 ^\circ$ ensures off-axis recording conditions. The temporal modulation of these fringes is controlled by the optical frequency shift $\Delta \omega$ between the reference and the illumination beam. In all the reported experiments, sets of 32 consecutive interferograms \{$I_1$, \ldots, $I_{32}$\} are recorded for each detuning frequency $\Delta \omega$, for signal averaging purposes. The measurement time of one frequency component is $32/12 = 2.7 \, \rm s$.

\section{In vitro fluid flow imaging}\label{sect_InVitro}

\begin{figure}[t]
\centering
\includegraphics[width = \linewidth]{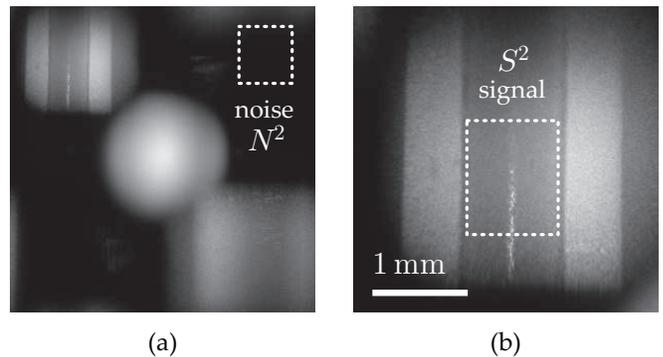}
\caption{(a) Squared amplitude hologram of the tube used for in vitro fluid flow calibration measurements. The image of the tube is in the upper left corner. The noise component $N^2$ (Eq.~\ref{eq_noise}) is averaged spatially in the upper right corner. The twin-image appears in the lower right corner. Self-beating interferometric contributions are gathered in the center of the hologram. (b) zoom in the region of the object. The signal component $S^2$ (Eq.~\ref{eq_2PhaseDemodulation}) is spatially-averaged in the intralipid flow region to form Doppler line shapes reported in Fig.~\ref{fig_TubeSpectra}.}\label{fig_TubeHolograms}
\end{figure}

\begin{figure}[t]
\centering
\includegraphics[width = \linewidth]{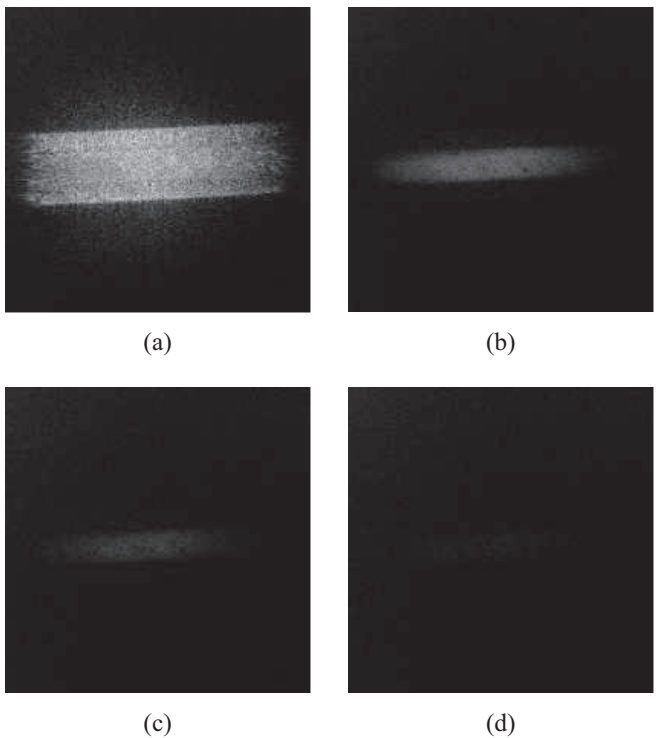}
\caption{Doppler images of a flowing intralipid suspension at velocity 0.1 mm/s in a transparent tube at four detuning frequencies $\Delta\omega/(2\pi)$ : 0 Hz (a), 200 Hz (b), 1 kHz (c), 2 kHz (d).}\label{fig_TubeDopplerMaps}
\end{figure}

From now on, throughout sections~\ref{sect_InVitro},~\ref{sect_DopplerBrain}, ~\ref{sect_DopplerEye}), the optical probe field $E(t)$ is described by a stochastic, random variable, as opposed to the model used to derive the temporal filter of the detection in section~\ref{sect_NarrowbandHeterodyneDetection}. The power spectrum is related to the optical field autocorrelation degree $g_1(t)$. A quantitative derivation of local decay rates of this function can be made from diffusing-wave spectroscopy formalism in backscattering configuration for a semi-infinite medium~\cite{Pine1988, RovatiCattini2007, LinHe2012, Cattini2013}
\begin{equation}\label{eq_g1_Delta_r_2}
g_{1} (t) \sim {\exp} \left( - \gamma \sqrt{k^{2} \left< \Delta r^{2}(t) \right>} \right)
\end{equation}
where $\left< \Delta r^{2}(t) \right>$ is the mean-square displacement of local light scattering particles, $\gamma$ is a fitting parameter, and $k$ is the optical wave number. Under the assumption that the optical field undergoes a stationary random phase fluctuations, according to the Wiener–Khinchin theorem~\cite{Wiener1930, bk_Saleh, bk_Goodman_stat_1985, Kumar2006}, the first-order Doppler spectrum $s_1$ and the field autocorrelation function $g_1$ form a Fourier pair
\begin{equation}\label{eq_s1_definition}
s_1(\omega) = \int _{- \infty} ^{+ \infty} g_1(\tau) \exp\left(-i \omega \tau \right)  {\rm d} \tau
\end{equation}
Detuning the LO frequency of the reference beam of the Mach-Zehnder interferometer by $\Delta \omega$ permits robust narrowband heterodyne imaging of a component of the first-order RF spectrum of the probe field, by forming the quantity $S^2/N^2\propto s_1(\Delta \omega)$ (Eq.\ref{eq_SNR}). In practice, for a given detuning frequency $\Delta \omega$, 32 consecutive interferograms are recorded \{$I_1$, ..., $I_{32}$\}. Off-axis hologram rendering in the image plane is performed with discrete Fresnel transformation~\cite{SchnarsJuptner1994, Schnars2002, KimYuMann2006, PicartLeval2008, VerrierAtlan2011}. The difference of consecutive off-axis holograms $H_{n+1}-H_{n}$ is then calculated in order to form the temporally-demodulated signal $S^2$ (Eq.~\ref{eq_2PhaseDemodulation}). This signal is temporally averaged over 32 raw frames. The noise component $N^2$ (Eq.~\ref{eq_noise}) is temporally averaged in the same manner. Moreover, the quantity $N^2$ is spatially averaged in the highlighted region of Fig.~\ref{fig_TubeHolograms}(a), and for the measurement of the Doppler lineshapes, the quantity $S^2$ is spatially-averaged in the intralipid flow region highlighted in Fig.~\ref{fig_TubeHolograms}(b). Representative Doppler maps at several detuning frequencies (0 Hz, 200 Hz, 1 kHz, and 2 kHz) for a flow speed of 0.1 mm/s are reported in Fig.~\ref{fig_TubeDopplerMaps}.


\subsection{Brownian motion of scatterers}

In the case of a suspension of diffusing particles for which the mean-square displacement is $\left< \Delta r^{2}(t) \right> = 6 D t$ , the optical field autocorrelation function can be written as~\cite{BernePecora2000} 
\begin{equation}\label{eq_g1_Brown}
g_{1} (t) \sim {\exp} \left( - \gamma \sqrt{6 t/\tau_0} \right)
\end{equation}
where $\tau_0 = (k^{2} D)^{-1}$, the parameter $\gamma $ depends on the configuration and $D$ is the spatial diffusion coefficient of a light-scattering particle (diffusivity). For particles in brownian motion, the theoretical first-order optical power spectral density (reported in Fig.~\ref{fig_TubeSpectra}, for null flow $v_{\rm F} = 0$) is calculated numerically from Eq.~\ref{eq_g1_Brown} and Eq.~\ref{eq_s1_definition}.

\subsection{Convective motion of scatterers}\label{sect_ConvectiveMotion}

Particles in deterministic (convective) motion with a quadratic mean velocity $v_{\rm F}$ undergo a local mean-square displacement 
\begin{equation}\label{eq_MSD_Convection}
\left< \Delta r^{2}(t) \right> = v_{\rm F}^{2}t^{2}
\end{equation}
The temporal autocorrelation function of the optical field backscattered from a suspension of flowing particles takes the form of an exponential decay; it scales as~\cite{WuPine1990}
\begin{equation}\label{eq_g1_Deterministic}
g_{1} (t) \sim {\exp} \left( - |t| / \tau \right)
\end{equation}
where $\tau = (\gamma k v_{\rm F})^{-1}$. For these particles in deterministic motion, the normalized first-order power spectral density of optical fluctuations is  
\begin{equation}\label{eq_s1_Deterministic}
s_1(\omega) \sim \frac{1}{1 + \omega^{2} / \omega_{\rm F}^{2}}
\end{equation}
The half-width at half-maximum $\omega_{\rm F}$ of this Lorentzian line shape is proportional to the local quadratic mean velocity $v_{\rm F}$
\begin{equation}\label{eq_OmegaFlow}
\omega_{\rm F} = \gamma k v_{\rm F}
\end{equation}
This quantity is the decay rate of $g_1$ (Eq.~\ref{eq_g1_Deterministic}). This line shape is the model against which the experimental Doppler measures are fitted, in vitro and in vivo, reported in Fig.~\ref{fig_TubeSpectra} (for $v_{\rm F}\neq 0$) and Fig.~\ref{fig_BrainSpectra}.

\subsection{Measurement of Doppler line shapes}

For the detected light composed of dynamic (subscript $_{\rm D}$) and static (subscript $_{\rm S}$) components ${\cal E} = {\cal E}_{\rm D} + {\cal E}_{\rm S}$ it was shown~\cite{BoasYodh1997, ZakharovVolker2006} that the autocorrelation function of the field will take the following form 
\begin{equation}\label{eq_g1_DynamicAndStatic}
g(t) = \rho |g_{1} (t)| + (1 - \rho)
\end{equation}
where $\rho = \left| {\cal E}_{\rm D} \right|^2 / (\left| {\cal E}_{\rm D} \right|^2 + \left| {\cal E}_{\rm S} \right|^2)$ characterizes the relative weight of the dynamic part of the detected light intensity, referred to as pedestal wave~\cite{Ogiwara1979, Agrawal1981}. In that case, the holographic measurement of the local RF spectrum of light, defined in Eq.~\ref{eq_SNR}, should exhibit the presence of the apparatus function $B$ (Eq.~\ref{eq_fPSF}) in the non-shifted light component. At a given reconstructed pixel, at a detuning $\Delta\omega$ of the local oscillator, measurements can be described by the following expression
\begin{equation}\label{eq_SNR_vs_s1}
S^2 / N^2 \propto \rho s_{1} (\Delta\omega) + (1-\rho) B(\Delta\omega)
\end{equation}

\subsection{Experimental measurements of in vitro flows}

\begin{figure}[t]
\centering
\includegraphics[width = \linewidth]{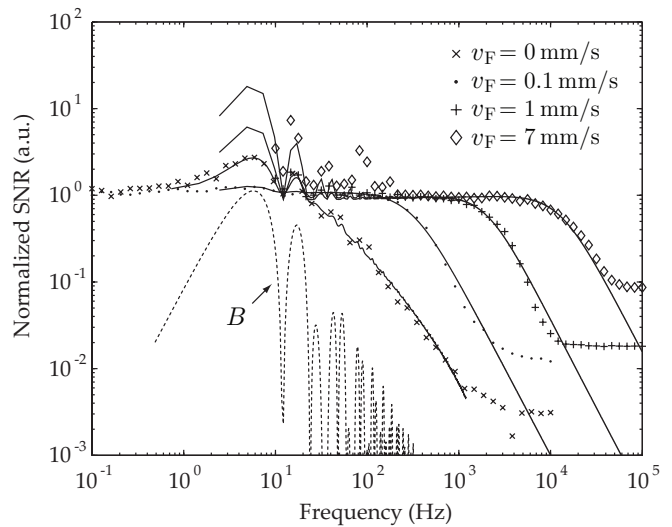}
\caption{Normalized spectral lines (a.u., symbols) versus detuning frequency, in Hz. The average perfusion velocity in the tube ranges from 0 mm/s to 7 mm/s (symbols). The measured lines are fitted against Eq.~\ref{eq_SNR_vs_s1} for brownian (Eq.~\ref{eq_g1_Brown}) and convective (Eq.~\ref{eq_s1_Deterministic}) motion of the scatterers, plotted as continuous black lines. The apparatus lineshape Eq.~\ref{eq_fPSF} is plotted as a dotted line.}\label{fig_TubeSpectra}
\end{figure}

\begin{figure}[t]
\centering
\includegraphics[width = \linewidth]{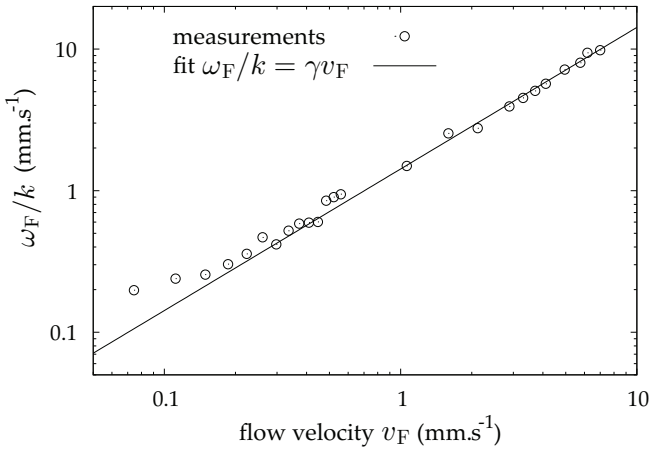}
\caption{Doppler broadening $\omega_{\rm F}/k$ (${\rm mm.s}^{-1}$) versus average perfusion quadratic velocity $v_{\rm F}$ (${\rm mm.s}^{-1}$). In vitro measurements of Doppler linewidth are in agreement with the model of convective motion (Eq.~\ref{eq_s1_Deterministic}) within a velocity range of 0.1 mm/s to 10 mm/s.}\label{fig_DopplerWidthVersusFlow}
\end{figure}

The setup for in vitro flow assessment consists of a tube of 1 mm diameter in which a 1 part in 10 dilute solution of 10\% Intralipid in water is injected with a known average velocity by a calibrated syringe pump. The range of velocities studied with this set up varies from $0 \, \rm mm.s^{-1}$ to $10 \, \rm mm.s^{-1}$. The experimental Doppler broadened lines measured for known velocities $v_{\rm F}$ were fitted against expression~\ref{eq_SNR_vs_s1}. The results for RF spectra obtained for 3 velocities (symbols) are shown on Fig.~\ref{fig_TubeSpectra} (black lines). The variation of the half width of the Doppler line $\omega_{\rm F}$ with the flow velocity $v_{\rm F}$ is investigated. We plotted $\omega_{\rm F} / k$ against $v_{\rm F}$ in Fig.~\ref{fig_DopplerWidthVersusFlow} in order to verify experimentally whether the relation $\omega_{\rm F} / k = \gamma v_{\rm F}$ holds. This plot shows that the Doppler width scales linearly with the flow velocity for flows in the 100 $\mu$m/s to 10 mm/s  range, which is in agreement with previously reported results, in vitro~\cite{Rousseau1971, Estes1971, Singh1992}. The coefficient $\gamma$ evaluated from this procedure is  
\begin{equation}\label{eq_gammaValue}
\gamma = 1.42
\end{equation}
which is consistent with experimental diffusing wave spectroscopy results in semi-infinite scattering media~\cite{Pine1988}. This value of $\gamma$ is used to calculate flow maps in section~\ref{sect_DopplerBrain} and section~\ref{sect_DopplerEye}. For flow velocities below 100 microns per second, experimentally measured Doppler widths are systematically larger than $k \gamma v_{\rm F}$,  which can be explained by the fact that the dominating motion of the scatterers is diffusive. This is confirmed by fitting the measured spectrum of light scattered from particles undergoing only brownian motion with the discrete Fourier transform of Eq.~\ref{eq_g1_Brown}, which yields the Doppler line for null flow velocity reported in Fig.~\ref{fig_TubeSpectra}. A spatial diffusivity $D = k_{\rm B} T / (6\pi \mu r) = 1/(k^{2}\tau_0) \sim 8\times 10^{-13} \, \rm m^{2}.s^{-1}$ is found for $\gamma = 1.42$. It corresponds to a homogenous suspension of oil droplets of radius $r = 311 \, \rm nm$, in water of dynamic viscosity $\mu = 0.89 \times 10^{-3} \, \rm kg.m^{-1}.s^{-1}$. The effect of residual motion on Doppler flow measurements is referred to as the biological zero problem~\cite{ZhongSeifalian1998, KernickTooke1999}.

\section{Imaging of microvascular cerebral blood flow}\label{sect_DopplerBrain}

Cerebral blood flow imaging in the mouse brain was conducted in strict compliance with approved institutional protocols and in accordance with the provisions for animal care and use described in the European Communities Council directive of 24 November 1986 (86-16-09/EEC). Two 26g C57BL6J mouses were anesthetized with urethane (1.75 mg/g). Paw withdrawal, whisker movement, and eyeblink reflexes were suppressed. The head of the mouse was fixed by using a stereotaxic frame (Stoelting). The skin overlying the right cerebral hemisphere was removed and the bone gently cleaned. A 3 $\times$ 3 mm craniotomy was made above the primary somatosensory cortex. Extreme care was taken at all times not to damage the cerebral cortex, especially during the removal of the dura. Physiological Ringers solution containing (in mM): 135 NaCl, 5 KCl, 5 HEPES, 1.8 CaCl2, and 1 MgCl2, was used during the surgery to prevent the exposed cortical area from drying out. At the end of this surgical procedure, the cortical surface was covered with agarose (1 \% in Ringers), and a coverslip was placed on top.\\

\begin{figure}[t]
\centering
\includegraphics[width = \linewidth]{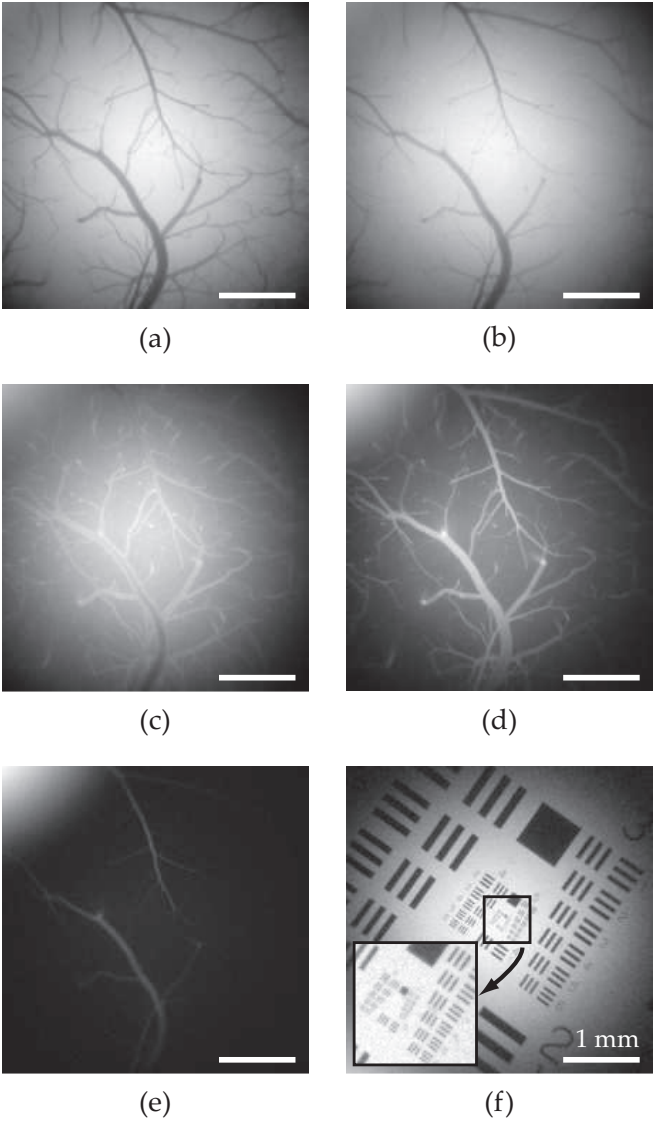}
\caption{Doppler images of the mouse cerebral cortex at different frequency shifts 9 Hz (a), 167 Hz (b), 1.7 kHz (c), 5.5 kHz (d). 17.5 kHz (e). The frequency sweep is reported in Media 1. Image of an USAF resolution target (f). Arbitrary logarithmic gray scale. Scale bar : 1 mm.}\label{fig_BrainDopplerMaps}
\end{figure}
\begin{figure}[t]
\centering
\includegraphics[width = \linewidth]{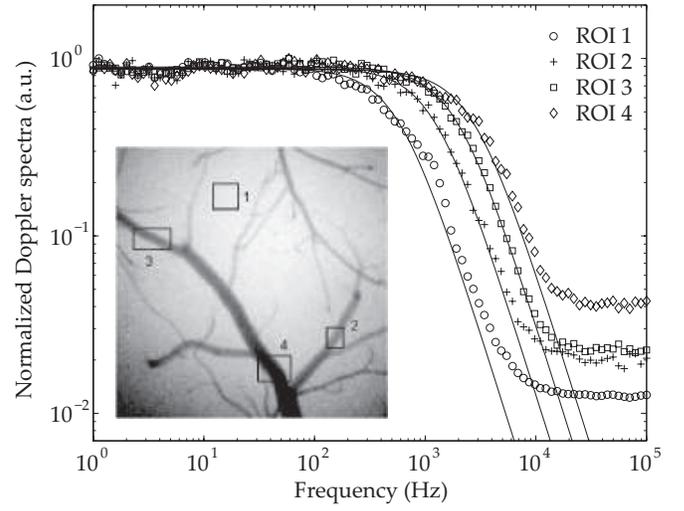}
\caption{ Normalized first-order power spectra averaged in the regions of interest (ROI) 1 to 4 (dots correspond to measurements, continuous lines are the result of fitting measured values against Eq.~\ref{eq_s1_Deterministic}). Insert: Doppler map displaying the 4 regions of interest.}\label{fig_BrainSpectra}
\end{figure}

Doppler maps were acquired for logarithmically-spaced LO detunings $\Delta\omega/(2\pi)$ from 1 Hz to 100 kHz, in order to observe the optical fluctuations due to the Doppler effect induced by blood flow. Fig.~\ref{fig_BrainDopplerMaps} shows representative Doppler images of the quantity $S^2/N^2$ (Eq.~\ref{eq_SNR}) in the cerebral cortex of a mouse, at five frequency shifts, 9 Hz (a), 167 Hz (b), 1.7 kHz (c), 5.5 kHz (d). 17.5 kHz (e). The speckle was attenuated by averaging Doppler maps over three consecutive frames. The comprehensive data obtained from the frequency sweep is reported in Media 1.  At low frequency (a), the background (parenchyma) exhibits a higher signal than the vessels. At higher frequency, the contrast is reversed and vessels exhibit a higher Doppler signal than the background. As the detuning frequency increases (d, e), only the main vessels are revealed. The inhomogeneity of the illumination can also be observed in those figures. First-order RF spectral line shapes, normalized by their value at low frequency are reported in Fig.~\ref{fig_BrainSpectra}. They are measured and averaged over four regions of interest, labeled from 1 to 4, ranked by increasing blood flow, according to the image shown in the insert. They show the features described herein, i.e. narrow broadenings for slow flows and broader line shapes for larger flows.\\

\begin{figure}[b]
\centering
\includegraphics[ width = 7 cm ]{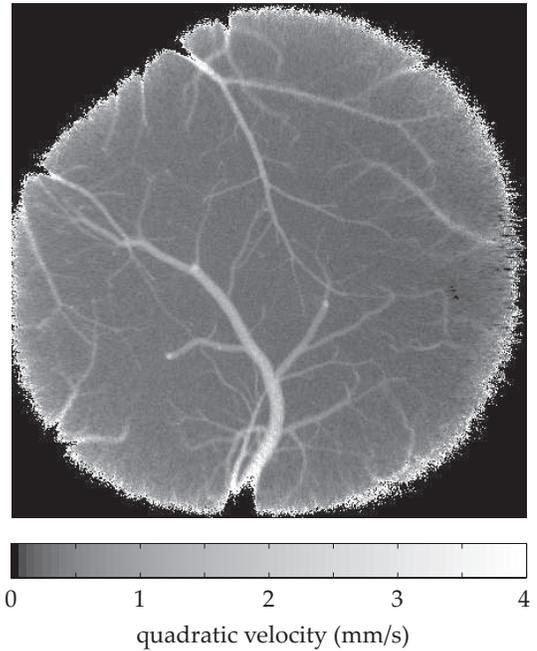}
\caption{Map of the local quadratic mean blood flow velocity $v_{\rm F}$ derived from the Doppler maps reported in Fig.~\ref{fig_BrainDopplerMaps}.}\label{fig_BrainFlowMap}
\end{figure}

We assume that bioflows under investigation lead to convective scatterers' motions much larger than the ones due to diffusive processes, so that the formalism of section~\ref{sect_ConvectiveMotion} is valid. if the local mean square displacement of an elementary scattering volume in the tissue scales up quadratically with time $t$, relationship~\ref{eq_MSD_Convection} holds and we can derive a quadradic mean flow map from a fitting procedure of the measured signal $S^2/N^2$ against Eq.~\ref{eq_SNR_vs_s1}. The temporal correlation function of the backscattered optical field $g_1(t)$ is of the form of Eq.~\ref{eq_g1_Deterministic}, and the first-order fluctuation spectrum has a Lorentzian lineshape, described by Eq.~\ref{eq_s1_Deterministic}.  The first-order power spectral density of the fluctuating field backscattered by the cortex agrees well with a Lorentzian line shape Eq.~\ref{eq_s1_Deterministic} whose half-width at half maximum is $\omega_{\rm F}= \gamma k v_{\rm F}$. The parameter $\gamma$ was assessed from the calibration procedure in vitro (Eq.~\ref{eq_gammaValue}), $k = 2 n \pi /\lambda$ and $n=1.39$ is the mean refractive index of the blood at our studying wavelength~\cite{Faber2004}, which is close to the refractive index of the brain, evaluated at 1.35 for $\lambda$=1.1 $\rm {\mu m}$~\cite{Binding2011}. A holographic image of a resolution target, acquired in the same experimental configuration, is reported in Fig.~\ref{fig_BrainDopplerMaps} (f), from which we can observe a lateral resolution limit of about 10 microns, which is compatible with the observation of the smallest vessels. Four spectra from the region of interest are reported in Fig.~\ref{fig_BrainSpectra}. They are normalized by their respective maximal value at low frequency. As we can see, the higher the blood flow velocity, the broader the spectra. Those spectra are fitted by Lorentzian lines from  Eq.~\ref{eq_s1_Deterministic} with a robust nonlinear least squares algorithm (MathWorks Matlab Curve Fitting Toolbox) and plotted as solid lines on the same graph. These theoretical lines are in good agreement with the measurements. A map of the quadratic mean blood flow velocity $v_{\rm F}$ can then be derived by fitting each pixel's Doppler line shape with the Lorentzian line of Eq.~\ref{eq_s1_Deterministic} to assess local values of $v_{\rm F}$. The result is reported in Fig.~\ref{fig_BrainFlowMap}. The inhomogeneity of the lighting seen in Fig.~\ref{fig_BrainDopplerMaps} is no longer observed. The resulting velocity map is thus independent on the local illumination level. The blood vessels are well resolved, even the smallest ones.

\section{Imaging of microvascular retinal blood flow}\label{sect_DopplerEye}

Retinal blood flow imaging was conducted in strict compliance with approved institutional protocols. Three adult rats were used for the preparations. Anesthesia was induced by intraperitoneal injection of 100 mg/kg ketamine and 25 mg/kg xylazine (both from Sigma-Aldrich). Topical tropicamide (CibaVision) was administered for pupil dilation. Each rat was placed on its side in a clay bed, their right eyes under the illumination beam. The head was supported so that the iris was perpendicular to the illumination axis. After administration of topical oxybuprocaine (CibaVision), a coverslip was applied on a ring surrounding the globe in order to compensate for the cornea curvature~\cite{SimonuttiPaquesSahel2010}. Methylcellulose (Goniosol) was applied as a contact medium.\\

\begin{figure}[t]
\centering
\includegraphics[width = \linewidth]{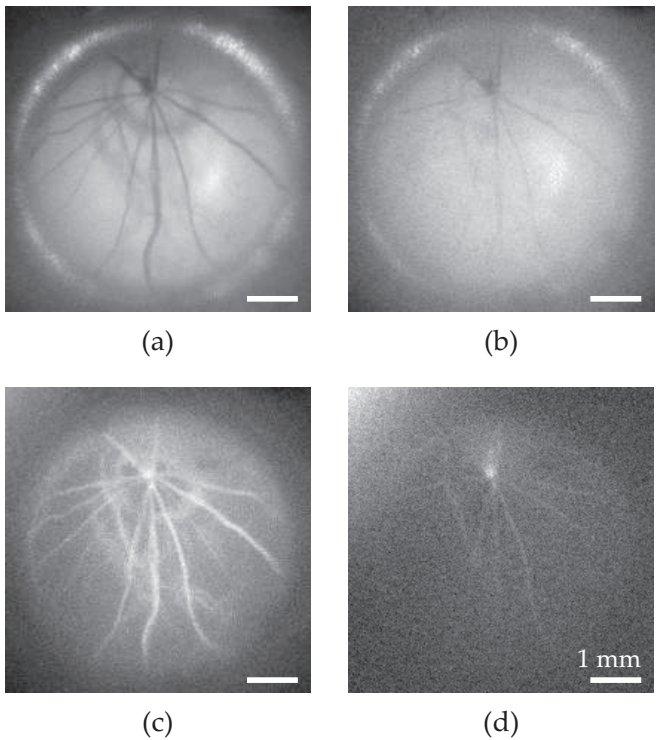}
\caption{Retinal images at four frequency shifts, 10 Hz (a), 513 Hz (b), 3.0 kHz (c), 6.1 kHz (d). The frequency sweep is reported in Media 2. Scale bar : 1 mm. }\label{fig_EyeFundusDopplerMaps}
\end{figure}
Doppler maps were acquired for logarithmically-spaced LO detunings $\Delta\omega/(2\pi)$ from 10 Hz to 63 kHz, in order to observe the optical fluctuations due to the Doppler effect induced by blood flow in the eye fundus. Fig.~\ref{fig_EyeFundusDopplerMaps} shows representative Doppler images of the quantity $S^2/N^2$ (Eq.~\ref{eq_SNR}) in the eye fundus of a mouse, at four frequency shifts : 10 Hz (a), 513 Hz (b), 3.0 kHz (c), 6.1 kHz (d). The speckle was attenuated by averaging Doppler maps over three consecutive frames. The comprehensive data obtained from the frequency sweep is reported in Media 2. The contrast reversal from low to high frequency is similar to the observations in the brain, reported in Fig.~\ref{fig_BrainDopplerMaps}. First-order RF spectral line shapes, normalized by their value at low frequency are reported in Fig.~\ref{fig_EyeFundusSpectra}. They are measured and averaged over five regions of interest, labeled from 1 to 5.

\subsection{Velocity composition}

The global motion of the eye has to be taken into account while considering the velocity of the blood flow. Let $\bf v_{\rm G}$ be the global in-plane velocity of the retina that defines eye movement~\cite{Riggs1954, Westheimer1975}. The scattering particles flowing in a vessel with an in-plane velocity $\bf v_{\rm F}$ have a total instant velocity
\begin{equation}\label{eq_VelocityComposition}
\bf v = v_{\rm F} + v_{\rm G}
\end{equation}
The local velocity outside a vessel is dominated by the global physiological motion at the quadratic mean velocity $v_{\rm G} = \sqrt{\left<{\bf v_{\rm G}}^2\right>}$, where $\left< \, \right>$ denotes averaging over the measurement time. We assume that the global physiological motion of the eye, induces a frequency broadening $\omega_{\rm G}/(2\pi)$ related to its quadratic mean velocity $v_{\rm G} = \omega_{\rm G} / (\gamma k)$, where $k = 2 \pi n / \lambda$ with $n$ the average refractive index of the eye fundus~\cite{GrievePaques2004}. The local retinal blood flow induces a frequency broadening $\omega_{\rm F}/(2\pi)$ related to the quadratic mean velocity $v_{\rm F} = \sqrt{\left<{\bf v_{\rm F}}^2\right>}$. In the neighborhood of a vessel, the local mean square displacement at time $t$ of an elementary scattering volume in the tissue is supposed to be, on average, over many realizations 
\begin{equation}\label{eq_MSD_composition}
\left< \Delta r ^2 (t) \right> \approx (v_{\rm F}^2 + v_{\rm G}^2) \, t^2
\end{equation}
under the hypothesis that the averaged cross-term $\left< 2 \bf v_{\rm F} \cdot v_{\rm G} \right>$ is zero, because the blood flow direction and the global motion direction are independent, and $v_{\rm G}$ is supposed to be a zero-mean random variable. Under these assumptions, the temporal correlation function of the optical field backscattered from the retina is the sum of two exponential decays
\begin{equation}\label{eq_g1_bi_exp}
g_1 (t) \sim (1-\rho) \exp \left( -t / \tau_{\rm G} \right) + \rho \exp \left( -t / \tau \right)
\end{equation}
where $\rho$ is the weight of light component frequency-shifted by the blood flow and $1-\rho$ is the weight of light affected by the global motion of the eye only. The decay times satisfy the relationships $\tau_{\rm G} = 1 / (\gamma k v_{\rm G})$, $\tau = 1/\sqrt{\tau_{\rm F}^{-2} + \tau_{\rm G}^{-2} }$ and $\tau_{\rm F} = 1 / (\gamma k v_{\rm F})$. Eq.~\ref{eq_g1_bi_exp} yields a first-order power spectral density of the form
\begin{equation}\label{eq_s1_retina}
s_{1} (\omega) \sim \frac{1-\rho}{ 1 + \omega^2/ \omega_{\rm G}^2} + \frac{\rho}{ 1 + \omega^2/ \ (\omega_{\rm G}^2 + \omega_{\rm F}^2)}
\end{equation}
where $\omega_{\rm G} = 1/\tau_{\rm G}$ and $\omega_{\rm F} = 1/ \tau_{\rm F}$, from which the quadratic mean blood flow velocity $v_{\rm F}$ can be derived with Eq.~\ref{eq_OmegaFlow}.

\subsection{Derivation of flow maps}
\begin{figure}[t]
\centering
\includegraphics[width = 7.5cm]{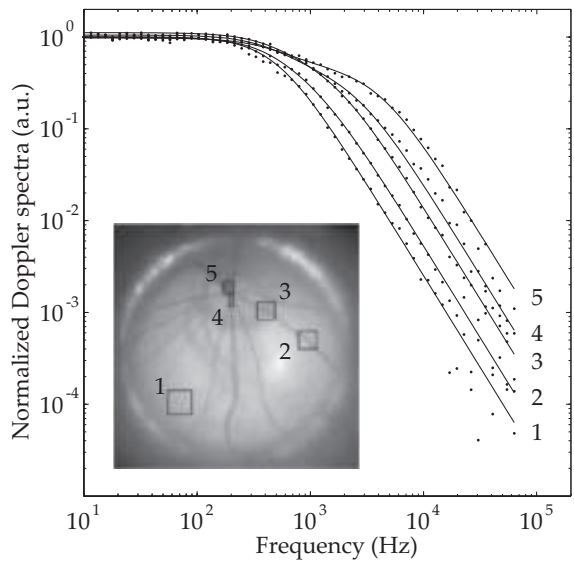}
\caption{Normalized first-order power spectra averaged in the regions of interest labeled from 1 to 5 (dots correspond to measurements, continuous lines are the result of the fitting procedure). Insert : Doppler map showing the 5 regions where spectra are measured.}\label{fig_EyeFundusSpectra}
\end{figure}
\begin{figure}[t]
\centering
\includegraphics[width = 7 cm]{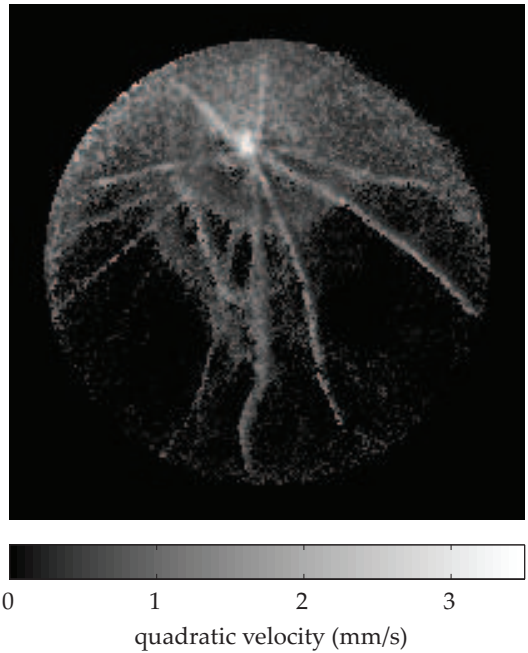}
\caption{Local quadratic mean velocity map in the eye fundus. Units : ${\rm mm.s^{-1}}$. This map was derived from Eq.~\ref{eq_OmegaFlow} and a fitting procedure of the Doppler maps against Eq.~\ref{eq_s1_retina} for $\rho >35\%$.}\label{fig_EyeFundusFlowMap}
\end{figure}
Experimental spectral lines reported in Fig.~\ref{fig_EyeFundusSpectra} were averaged over several regions of interest, labeled from 1 to 5, and were fitted by a least squares fitting algorithm with Eq.~\ref{eq_s1_retina}, with the robust nonlinear technique of MathWorks' Matlab Curve Fitting Toolbox. An estimation of the average Doppler broadening $\omega_{\rm F}$ at each point of the image is derived from the fitting parameters of Eq.~\ref{eq_s1_retina}. Once  $\omega_{\rm F}$ is assessed, an estimation of the local quadratic mean blood flow velocity $v_{\rm F}$ can be derived, besides a factor $\gamma$, via Eq.~\ref{eq_OmegaFlow}. The characteristic angular frequency $\omega_{\rm G}$ of the global physiological movements can be determined from a region showing no perfusion. In this case, only one Lorentzian line is needed to describe the first order optical power spectral density. A region is selected, labeled 1 in the insert of Fig.~\ref{fig_EyeFundusSpectra}. The Doppler line averaged in region 1 is shown in Fig.~\ref{fig_EyeFundusSpectra}. It leads to a global background Doppler broadening value of $\omega_{\rm G} / (2\pi) \sim $ 500 Hz, and to a corresponding quadratic mean velocity of $v_{\rm G} \sim $ 0.2 mm/s, for a value of $\gamma$ calibrated in vitro (Eq.~\ref{eq_gammaValue}). This velocity is the threshold value above which local flow quadratic mean velocities for retinal imaging in the reported experimental conditions can be assessed, for eye fundus imaging of anesthetized rats. The local quadratic mean blood flow velocity map derived from Eq.~\ref{eq_s1_retina} is shown in Fig.~\ref{fig_EyeFundusFlowMap}. Some artifacts can be seen in the surrounding medium while no vessels are seen. Therefore, the local relative weight of the flow component can be used to wash-out fitting artifacts and enhance local flow maps. The quantity $\rho$ can be assessed from the fitting procedure. For the rms velocity map shown on Fig.~\ref{fig_EyeFundusFlowMap}, only the velocity corresponding to a weight greater than 35\% is taken into account. Below the threshold value of 35\%, the fitting procedure has the tendency to derive aberrant velocity values from noise.

\section{Conclusions}

We demonstrated that holographic laser Doppler imaging has the potential to enable quantitative assessment of hemodynamic parameters. Cerebral and retinal blood flow was mapped in the superficial microvasculature of rodents. Frequency-scanned narrowband detection of Doppler components with a spectral resolution of a few Hertz was performed;  Doppler spectra at radiofrequencies up to 100 kHz for a radiation wavelength of 785 nm revealed contrasts of microvascular blood flow. We derived quantitative quadratic mean blood flow velocity maps by using the first-order power spectrum of optical fluctuations from a basic inverse-problem model involving the diffusing-wave spectroscopy formalism. In vitro validation of this method allowed steady-state rms assessment of fluid flow from $100\, \mu {\rm m}. {\rm s}^{-1}$ to $10\, {\rm mm}. {\rm s}^{-1}$. The low speed limit being due either to the dependence of the signal on dynamic background flows or thermal motion. The lateral spatial resolution of about 10 microns is compatible with the visualization of the smallest superficial vessels and arteries, but because of random scattering of light in tissue, flow direction is lost, and no depth sectioning is demonstrated here, in contrast to optical Doppler tomography~\cite{Wang2008}. Furthermore, strong hypotheses are made in the inverse problem formulation, which include multiple random light scattering in a semi-infinite medium, homogenous optical index of refraction, steady-state flows, uncorrelated velocities, and the presence of a fitting parameter.\\

Optical techniques provide suitable non-contact ways of obtaining superficial microvascular blood flow images at high resolution. They are good candidates for the development of robust quantitative, non-invasive, non-ionizing motion screening tools. The key benefits of holography with respect to the state-of-the-art optical schemes for blood flow imaging is its propensity to reveal Doppler contrasts of microvascular blood flow in low-light, which is a strong competitive advantage for retinal monitoring, and quantitative fluctuation spectra, from which local dynamical properties can be assessed. Furthermore, holography is suited to the design of robust microrheological imaging tools without any contrast agent, that could be used for clinical exploration of retinal blood flow. The major drawback of time-averaged holographic detection of blood flow is poor temporal resolution, due to sequential frequency-scanning of the Doppler spectrum. In particular, pulsatile flow could not be assessed. Potential ways of circumventing this issue may be to limit the acquisition to one Doppler component with a faster camera, or to have recourse to high-speed Fourier-transform Doppler imaging~\cite{SamsonVerpillat2011, SamsonAtlan2013} and/or logarithmic frequency chirps of the detection frequency. Also, local~\cite{Leutenegger2011} or even on-chip~\cite{LaforestDupret2013, HeNguyen2013} processing of the optical measurements with sensor arrays can be investigated. 

\section*{Acknowledgements}

We gratefully acknowledge support from Agence Nationale de la Recherche (ANR-09-JCJC-0113, ANR-11-EMMA-046), Fondation Pierre-Gilles de Gennes (FPGG014), r\'egion Ile-de-France (C'Nano, AIMA), the "investments for the future" program (LabEx WIFI: ANR-10-LABX-24, ANR-10-IDEX-0001-02 PSL*), and European Research Council (ERC Synergy HELMHOLTZ).


\end{document}